\documentclass[twocolumn,amsmath,amssymb,floatfix,prl,showpacs,footinbib]{revtex4}
\usepackage{amsmath,amssymb,natbib,bm,graphicx,times}
\usepackage[ansinew]{inputenc}

\usepackage{color}

\newcounter{fig}

\newcommand{\NN}{\mathbb{N}}
\newcommand{\ZZ}{\mathbb{Z}}

\newcommand{\abs}[1]{\left|#1\right|}

\newcommand{\bra}[1]{\left\langle \, #1 \,\right|}
\newcommand{\ket}[1]{\left| #1  \right\rangle}
\newcommand{\bket}[2]{\left\langle  #1 | #2  \right\rangle}
\newcommand{\boket}[3]{\langle #1 | #2 | #3 \rangle}

\newcommand{\be}{\begin{equation}}
\newcommand{\ee}{\end{equation}}

\begin{document}
\title{Charging effects in the inductively shunted Josephson junction}
\author{Jens Koch} 
\affiliation{Departments of Physics and Applied Physics, Yale University, New Haven, Connecticut 06520, USA}
\author{V.~Manucharyan} 
\affiliation{Departments of Physics and Applied Physics, Yale University, New Haven, Connecticut 06520, USA}
\author{M.~H.~Devoret} 
\affiliation{Departments of Physics and Applied Physics, Yale University, New Haven, Connecticut 06520, USA}
\author{L.~I.~Glazman}
\affiliation{Departments of Physics and Applied Physics, Yale University, New Haven, Connecticut 06520, USA}
\date{September 3, 2009}
\begin{abstract}
The choice of impedance used to shunt a Josephson junction determines if the charge transferred through the circuit is quantized: a  capacitive shunt renders the charge discrete, whereas an inductive shunt leads to continuous charge. This discrepancy leads to a paradox in the limit of large inductances $L$. We show that while the energy spectra of the capacitively and inductively shunted junction are vastly different, their high-frequency responses become identical for large $L$. Inductive shunting thus opens the possibility to observe charging effects unimpeded by charge noise.
\end{abstract}
\maketitle

\begin{figure}
	\centering
		\includegraphics[width=1.0\columnwidth]{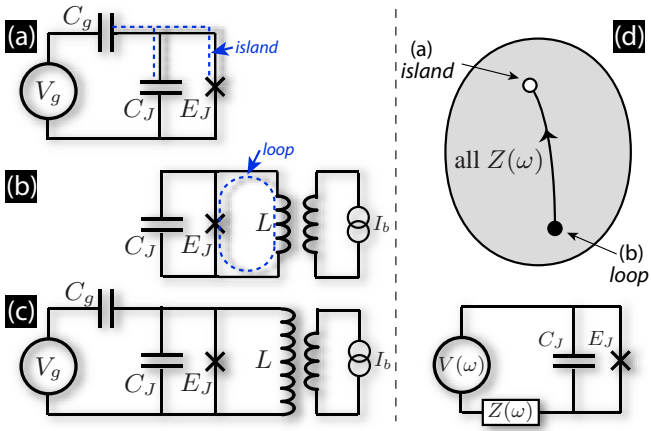}
	\caption{(color online) Examples of dispersively shunted Josephson junctions. (a) Island-based devices, like the Cooper pair box, are obtained by shunting a Josephson junction capacitively and are typically operated by coupling to charge. (b) Loop-based devices like the one-junction flux or phase qubit use an inductive shunt of the Josephson junction and are commmonly operated by coupling to flux. (c) For large inductances as in the fluxonium device, both charge and flux coupling play a role.
	(d) According to Th\'evenin's theorem, both circuits (a) and (b) represent specific realizations of a Josephson junction shunted by an impedance $Z(\omega)$. Taking the limit of large inductance should smoothly connect the  realization (c) to (a).\label{fig:figure1}}
\end{figure}

The Josephson junction plays a key role in superconducting devices, being the only nonlinear yet dissipationless element in a quantum circuit  \cite{devoret__2004}. The simplest examples of nonlinear quantum circuits are obtained by shunting a single Josephson junction by a purely dispersive impedance $Z(\omega)$. Following this prescription and using a capacitive shunt, one obtains the Cooper pair box (CPB) \cite{buttiker_1987,bouchiat_quantum_1998,nakamura_coherent_1999}, see Fig.\ \ref{fig:figure1}(a). Alternatively, using an inductive shunt, one obtains the single-junction flux qubit \cite{friedman_quantum_2000} and the phase qubit \cite{martinis_rabi_2002}, see Fig.\ \ref{fig:figure1}(b).

There is an important difference between these two examples: in the circuit of Fig.\ \ref{fig:figure1}(a), one of the terminals of the Josephson junction forms a superconducting island coupled only capacitively to the rest of the circuit, while the circuit of Fig.\ \ref{fig:figure1}(b) does not contain isolated pieces of superconductor. It is well-known that charges of two superconductors coupled by only a Josephson junction are quantized at the level of eigenvalues of the corresponding operators. Adding an inductive link between the superconductors destroys that quantization. 

Paradoxically, intuition suggests that charge quantization should manifest itself as long as the inductance remains sufficiently large. Specifically, the properties of the inductively shunted junction [Fig.\ \ref{fig:figure1}(c)] should  approach those of the CPB [Fig.\ \ref{fig:figure1}(a)] as the inductance $L$ is increased. Indeed, using Th\'evenin's theorem, the environment of the Josephson junction can formally be described by the series impedance $Z_\text{Th}=i\omega L(1-\omega^2 L C_g)^{-1}$ [Fig.\ \ref{fig:figure1}(d)], which converges to the CPB expression $Z_\text{CPB}=(i\omega C_g)^{-1}$ as $L\to\infty$.

The central question formulated and answered in this Letter is whether and in what sense the charge of finite-size
superconductors remains quantized in the presence of a material link between them. We find
the proper observable quantity that allows one to see the manifestation of charging effects and its evolution with the
inductance of the material link. In addition to the fundamental interest, our observation has a
practical value \cite{vlad}: excited states of the inductively shunted Josephson junction turn out to be robust
 with respect to charge fluctuations while being strongly anharmonic, which is an advantage over earlier existing superconducting devices.

We start with the Hamiltonian describing the inductively shunted junction as shown in Fig.\ \ref{fig:figure1}(c),
\be\label{hams}
H=4E_C(n-n_g)^2 - E_J\cos \varphi +\frac{1}{2}E_L(\varphi+2\pi\Phi/\Phi_0)^2.
\ee
Here, the canonically conjugate operators $n=q/2e$ and $\varphi$ correspond to the junction capacitor charge (in units of Cooper pairs) and the superconducting phase difference across the junction, and they obey the relation $[\varphi,n]=i$. The three energy scales entering $H$ are the (single-electron) charging energy $E_C=e^2/2(C_J+C_g)$, the Josephson energy $E_J$, and the inductive energy $E_L=(\Phi_0/2\pi)^2/L$, defined in terms of the flux quantum $\Phi_0=h/2e$. The external magnetic flux through the loop (generated by the bias current $I_b$) is denoted by $\Phi$, and the effect of the gate voltage is expressed in the offset charge $n_g$. We note that the spectrum is invariant with respect to static offset charge, which can be confirmed by applying a gauge transformation $\bar\psi(\varphi)=e^{i n_g \varphi}\psi(\varphi)$. 

In spite of the familiarity of the Hamiltonian \eqref{hams}, its physics beyond the flux and phase qubit regimes ($E_C\ll E_L\alt E_J$) has remained largely unexplored, with the notable exception of the work by A.\ Kitaev \cite{kitaev_2008}. We focus on the regime of large inductances where $E_L$ represents the smallest energy scale, $E_L\ll E_C,E_J$.  In the first step, we show that large-inductance limit does \emph{not} smoothly transform the energy spectrum of the inductively shunted device into that of the CPB. Ultimately, the reason for the subtlety of this limit stems from the opposed symmetries of the two systems: while the CPB has a strictly periodic potential when written in the phase basis, the inductive shunt always breaks this periodicity. Equivalently, the discrepancy may be formulated in the charge basis: due to the presence of an island in the CPB, charge on the corresponding node is quantized in terms of Cooper pairs, whereas the inductive shunt eliminates the island and renders the charge continuous. 

The key to our analysis  consists of transforming the Hamiltonian \eqref{hams} into the basis of Bloch waves $\{\ket{s,p}\}$, where $s\in\NN$ is the band index and $p\in[0,1)$ the quasimomentum. These states diagonalize the CPB part [comprising the terms $\sim E_C$ and $E_J$] of the Hamiltonian \eqref{hams}, $H_0\ket{s,p}=\varepsilon_s(p)\ket{s,p}$, where the eigenenergies represent the usual CPB bands. For the transformation of the inductive term, we employ the standard relation $\varphi=id/dp+\Omega$ \cite{lifshitz_statistical_1980}, where the Hermitian operator $\Omega$ causes interband coupling and is defined via
\begin{align}
\boket{ps}{\Omega}{p's'} &= \delta(p-p')\frac{i}{2\pi}\int_0^{2\pi} d\varphi\, u_{ps}^*(\varphi) \frac{du_{ps'}}{dp}(\varphi).
\end{align}
Here, $u_{ps}$ is the Bloch amplitude, $\bket{\varphi}{s,p}=e^{i p\varphi}u_{s,p}(\varphi)$. For large $E_J/E_C$ and low-lying bands $s$, $s'$ we find
$\Omega_{ss'}(p)\approx(2E_C/E_J)^{1/4}(\sqrt{s}\delta_{s,s'+1}+\sqrt{s'}\delta_{s,s'-1})/2\pi$
so that interband coupling can be neglected. This simplifies the problem dramatically: the Hamiltonian becomes block-diagonal and separates into effective Hamiltonians
\be\label{hs}
H^{(s)}=\frac{E_L }{2}\left(i\frac{d}{dp}+\frac{2\pi\Phi}{\Phi_0}\right)^2 + \varepsilon_s(p)
\ee
for the low-lying bands. The spectra, obtained from $H^{(s)}$ with periodic boundary conditions in $p$, have to be overlaid. We note that Eq.~\eqref{hs} is structurally identical to the CPB Hamiltonian, and in the following we will draw from the knowledge of the CPB in both charge and transmon regimes \cite{bouchiat_quantum_1998,koch_charge-insensitive_2007}.

\begin{figure}
	\centering
		\includegraphics[width=0.9\columnwidth]{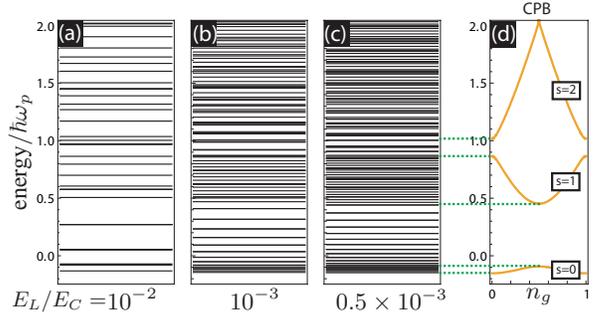}
	\caption{(color online). Energy spectra for the inductively shunted Josephson junction at zero flux. The progression (a)--(c) shows energy spectra for $E_J/E_C=2.5$ and decreasing inductive energy $E_L$. Energies are given in units of the plasma oscillation frequency $\hbar\omega_p=\sqrt{8E_JE_C}$. For small $E_L$, the spectrum exhibits regions with significantly different level spacings. (d) Comparison with the CPB bands for the same $E_J/E_C$ ratio reveals that regions of small (large) level spacings coincide with band (gap) regions in the CPB.	
\label{fig:figure2}}
\end{figure}
According to Eq.\ \eqref{hs}, $E_L$ determines the kinetic energy while the CPB bands $\varepsilon_s(p)$ act as potentials for the inductively shunted device. The potentials are periodic and can be expressed as Fourier series 
$\varepsilon_s(p)=\sum_{\ell=0}^\infty \varepsilon_{s,\ell} \cos(2\pi \ell p)$, and approach simple sinusoids for large $E_J/E_C$. For small inductive energies, each low-lying CPB band supports two types of states: (i) \emph{metaplasmon states} with approximate level spacing $2\pi\sqrt{E_L\abs{\varepsilon_{s,1}}}$, which are bound states in the CPB potential, and are associated with large charge oscillation across the Josephson junction, and (ii) \emph{persistent-current states}, existing in the gaps of the ``bare'' ($E_L=0$) CPB device; these states are associated with substantial persistent currents circulating (at finite $E_L$) through the inductor.
As a result, the energy spectrum of the shunted junction separates into regions with small level spacings fitting into the bands of the CPB, and large level spacings in the regions of CPB gaps. This distinction grows stronger with the decrease of the ratio $E_L/E_C$, see Fig.\ \ref{fig:figure2}(a)-(c).

The nature of the persistent-current states can further be clarified by rewriting the Hamiltonian \eqref{hs} in the discrete local-minimum basis $\{\ket{m,s}\}$, which formally parallels the charge basis of the CPB, 
\begin{align}\label{quasic}
H^{(s)}=&\frac{(2\pi)^2}{2}E_L(m+\Phi/\Phi_0)^2\\\nonumber
&+\frac{1}{2}\sum_{\ell=0}^\infty  \sum_{m=-\infty}^\infty \varepsilon_{s,\ell} \bigg[ \ket{m,s}\bra{m+\ell,s}+\text{H.c.}\bigg].
\end{align}
Physically, $\ket{m,s}$ corresponds to a state localized in the $m$-th local minimum of the phase-basis potential, carrying a persistent current $I_m^{(s)}=  (m\Phi_0 + \Phi)/L$. From Eq.~\eqref{quasic}, the energies of persistent-current states are found to be 
\be\label{pcstates}
E_{m}^{(s)}\approx \varepsilon_{s,0} + 2\pi^2 E_L (m+\Phi/\Phi_0)^2,
\ee
 where $m\in\ZZ$ has sufficiently large mod\-ulus so that $E_{m}^{(s)}>\max_p \varepsilon_s(p)$. Thus, to lowest order in the inter-well tunneling, the persistent-current states are doubly degenerate at zero flux, and each degenerate subspace is spanned by the two counter-propagating states with currents $I_{\pm m}^{(s)}\approx \pm m \Phi_0/L$ around the superconducting loop. At higher order in the inter-well tunneling, this degeneracy is lifted and time-reversal symmetry, which requires $I=0$ for vanishing magnetic flux, is restored. The resulting avoided crossings are extremely small. Specifically, for a purely sinusoidal $s=0$ band, i.e.\ $\varepsilon_{0,\ell}=0$ for $\ell>1$, the splitting between $\ket{m,s=0}$ and $\ket{-m,s=0}$ is generated by tunneling through the $2m$ potential barriers separating the two states, and scales as
\be
\delta E_m\approx \varepsilon_{0,1}\left[ \frac{\varepsilon_{0,1}}{(2\pi)^2E_L}\right]^{2m-1}\frac{1}{[(2m-1)!]^{2}}.
\ee

\begin{figure}
	\centering
		\includegraphics[width=1.0\columnwidth]{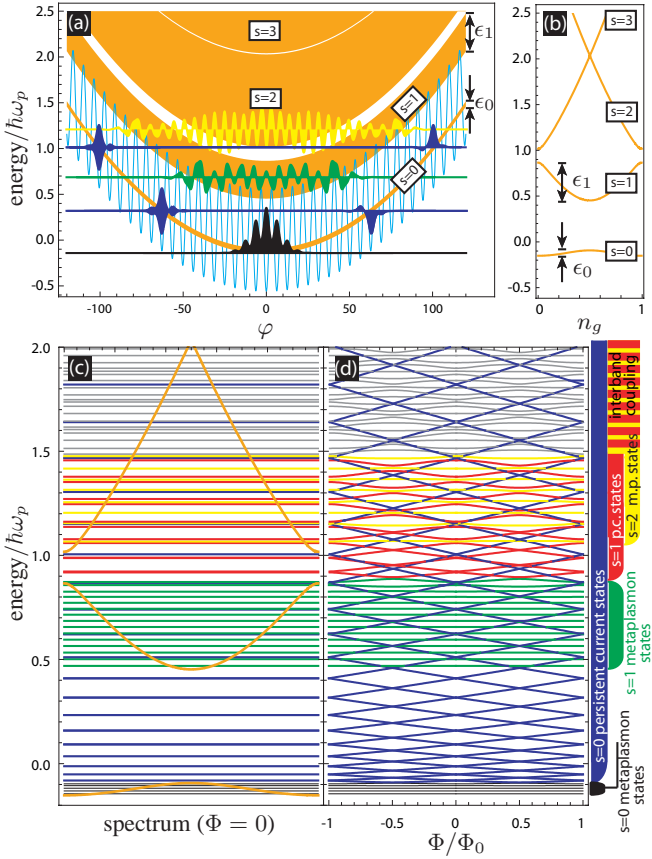}
	\caption{(color). Metaplasmon and persistent-current states for large inductances. Panel (a) shows the potential in $\varphi$-space (cyan), the corresponding bent CPB bands (orange), and examples of wavefunctions at zero flux, from bottom to top: black--ground-state wavefunction (lowest $s$=0 metaplasmon state), blue--$s$=0 persistent-current states, green--lowest $s$=1 metaplasmon state, and yellow--$s$=2 metaplasmon state. Localization in $\varphi$-space differs characteristically, with metaplasmon states being centered at $\varphi=0$ and delocalized, while persistent-current states are symmetric and antisymmetric superpositions of wavefunctions localized towards the edges of the parabolically deformed bands. (b) Corresponding CPB spectrum for comparison. Using the same color coding, (c) and (d) show the spectrum overlayed by the CPB bands, and the flux dependence of energy levels, respectively. Metaplasmon states (independent of flux) and persistent-current states (sensitive to flux) are easily distinguished. For the parameters chosen here, $E_J/E_C=2.5$ and  $E_L/E_C=10^{-3}$, interband coupling becomes significant at higher energies (levels in gray), leading to avoided crossings between $s=1$ persistent-current states (in red) and $s=2$ metaplasmon states. \label{fig:figure3}}
\end{figure}

Metaplasmon and persistent-current states also differ in their localization and respective flux dependence. As evident from Eq.\ \eqref{quasic} and illustrated in Fig.\ \ref{fig:figure3}(a), metaplasmon states are typically delocalized across several wells around $\varphi=0$, whereas persistent-current tend to localize in $\varphi$-space in the region of the corresponding parabolically deformed band.  For the bound metaplasmon states, the flux dependence is exponentially suppressed; specifically, variations in the metaplasmon states of band $s$ scale as $\sim\exp[-4\sqrt{\abs{\epsilon_{s,0}}/E_L}/\pi]$. By contrast, Eq.\ \eqref{pcstates} predicts that persistent-current states are strongly sensitive to magnetic flux. For the lowest bands, these predictions are confirmed by results from numerical diagonalization, see Fig.\ \ref{fig:figure3}(d). Persistent-current states exist above the lowest CPB band and overlap with $s>0$ metaplasmon states. At higher energies, avoided crossings between metaplasmon and persistent-current states become more significant, as interband coupling increases for higher bands.

\begin{figure*}
	\centering
		\includegraphics[width=0.9\textwidth]{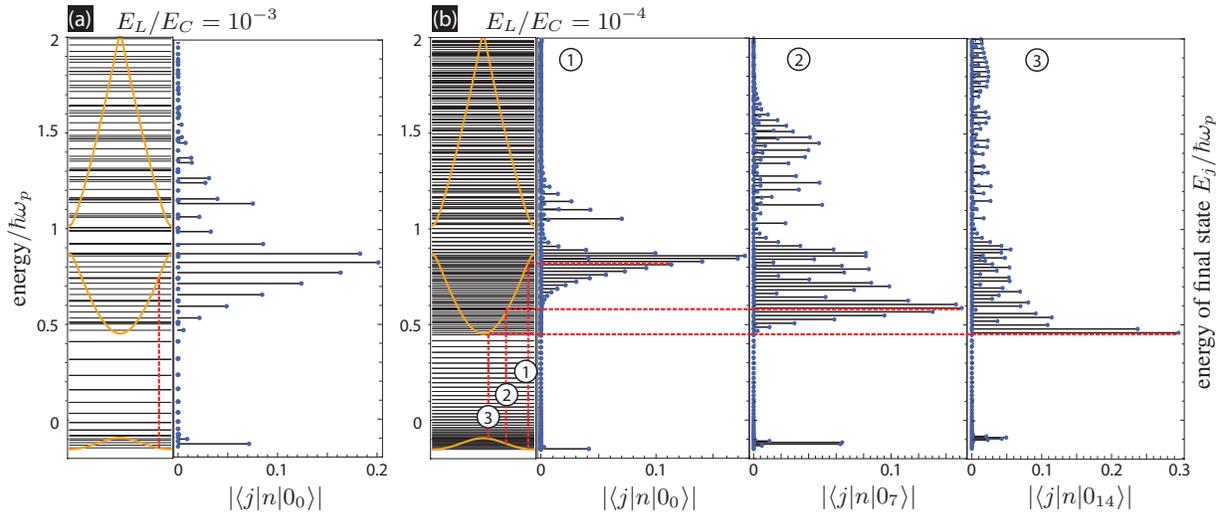}
	\caption{(color online). Matrix elements relevant for excitation when coupling the junction charge to an ac voltage. The magnitude of the matrix element for a transition to final state $\ket{j}$ is shown as a function of the final state energy $E_j$, in (a) for $E_L/E_C=10^{-3}$ starting in the ground state, and in (b) for $E_L/E_C=10^{-4}$ for several initial states. The prominent gap in the matrix elements (right panels) demonstrates the strong suppression of transitions from metaplasmon states to persistent-current states. For transitions among metaplasmon states, the matrix elements reach a distinct maximum for the transition occuring in the corresponding CPB without the inductive shunt (vertical dashed lines in left panels).  Lowering $E_L$ makes the peak narrower, and eventually leads to a selection rule excluding all but the CPB-allowed transitions. \label{fig:figure4}}
\end{figure*}

The previous discussion underlines the differences between the spectra of the CPB and the inductively shunted Josephson junction: the presence of the inductor introduces new levels located in the band gaps of the CPB, and these levels do not disappear in the limit of small $E_L$. However, we will now argue that in the limit of large inductance, the ac properties of the shunted device approach those of the CPB. To be specific, we consider the realistic \cite{koch_charge-insensitive_2007} coupling of the junction charge $n$ to an ac voltage potential, which modifies the offset charge term in Eq.\ (1), $n_g\to n_g+ C_g V_\text{rms}(a+a^\dag)/2e$. Here, $a$, $a^\dag$ annilate or create a photon in the microwave field mode, $V_\text{rms}$ denotes its root-mean square voltage. Transitions are then induced by tuning the field into resonance, and the transition strength depends directly on the charge matrix elements $\boket{f}{n}{i}$, where $i$, $f$ denote the initial and final states of the transition. At low energies where interband coupling is negligible, these states can be identified as eigenstates of Eq.\ \eqref{hs},
and we will write, e.g., $\ket{i}=\ket{s_\nu}$, thus denoting the $\nu$-th eigenstate belonging to band $s$. We find
\begin{align}
\boket{s'_{\nu'}}{n}{s_\nu}=&i (E_J/2E_C)^{1/4}(\sqrt{s}\delta_{s,s'+1}-\sqrt{s'}\delta_{s,s'-1})\nonumber\\
&\times\int_{-1/2}^{1/2} dp\,\chi_{s'\nu'}^*(p)\chi_{s\nu}(p).
\end{align}
The last integral involves the quasimomentum representation $\chi_{s\nu}(p)\equiv\bket{sp}{s_\nu}$ of eigenstates, and is amenable to evaluation within the WKB approximation. In complete analogy to the Franck-Condon principle, the matrix elements peak when the transition between the corresponding classical turning points occurs vertically, see Fig.\ \ref{fig:figure4}. For the metaplasmon transitions from $s=0$ to $1$ we find that below this optimum  the matrix elements decay exponentially with $\sim\exp[-\sqrt{2\abs{\varepsilon_{0,1}}{E_L}}(\eta/\abs{\varepsilon_{1,1}})^{3/2}/2\pi]$, whereas for energies above, the decay is oscillatory and follows an overall power-law decay $\sim[E_L^3/(\abs{\varepsilon_{11}\varepsilon_{01}}\eta)]^{1/4}$, where $\eta$ is the energy deviation from the optimum. As a result, in the limit of large inductances the matrix elements select the CPB-allowed transitions, thus producing pronounced charging effects.


We now discuss the charge-noise sensitivity of the inductively shunted junction, and show that the charging effects revealed in the ac response remain largely unspoiled by $1/f$ charge noise.
The origin of the reduced sensitivity to charge noise is identical to that of the the single-junction flux qubit and the flux-biased phase qubit \cite{simmonds_decoherence_2004}. Closing the Josephson junction by an inductive load allows one to evade the most severe problems caused by $1/f$ charge noise. Specifically, the vanishing dc impedance of the inductive element renders the energy spectrum independent of any constant offset charge, and transforms the $1/f$ charge noise into a relatively benign ``$f$-noise". 

To confirm this, consider the effects of a fluctuating offset charge $n_g(t)$ in the Hamiltonian \eqref{hams}. Again, we apply a gauge transformation $\bar\psi(\varphi)=e^{i n_g \varphi}\psi(\varphi)$. Taking into account the time dependence of $n_g$, this results in a Hamiltonian $\bar H$, obtained from $H$ by eliminating the offset charge $n_g$ from the charging term and appending a term $-\hbar \dot n_g\varphi$. Thus, the energy spectrum only depends on the time derivative of the offset charge, in a fashion similar to the dependence on flux $\Phi$, see Eq.\ \eqref{hams} \footnote{We emphasize that this reasoning does \emph{not} apply to the Cooper pair box, where the gauge transformation  affects the boundary conditions.}. The relevant noise spectrum for $\dot n_g$ is $S_{\dot{n}_g}(\omega)=\omega^2 S_{n_g}(\omega)$, 
such that Gaussian charge noise with a $1/f$ spectrum, $S_{n_g}(\omega)=A/\abs{\omega}$, is transformed into benign ``$f$-noise". Assuming an ultraviolet cutoff of the  $1/f$ spectrum at frequency $1/\tau_u$, the residual effect of $\dot n_g$ noise away from flux sweet spots and for long times $t\gg\tau_u$ is controlled by the behavior of the integral $g(t)=(\partial \omega_{ij}/\partial \dot n_g)^2\int_{1/t}^{1/\tau_u}d\omega\,S_{n_g}(\omega)$, such that off-diagonal elements of the density matrix decay as $\rho_{ij}\sim\exp[-g(t)]$. For large $t$, the integral $g(t)$ is logarithmically divergent,  and thus yields a slow power-law decay with exponent proportional to $(\partial\omega_{ij}/\partial \Phi)^2$. Remarkably, the \emph{flux} sweet spots thus become first-order insensitive to charge noise, and metaplasmon states with their suppressed flux dependence are expected to be especially well-protected.
The slow power-law decay should be contrasted with the rapid loss of coherences in the absence of an inductive shunt, which follows an exponential decay.

To summarize, the physics of the inductively shunted Josephson junction in the large-$L$ limit is distinct from the ones accessed by flux and phase qubits. Two different types of states, metaplasmon and persistent current states, with distinct level spacings and magnetic flux dependence dominate the spectrum at low energies. While the spectrum of the junction with large inductive shunt also differs from that of the CPB, we have demonstrated that the ac response due to charge coupling approaches the well-known response of the CPB, thus resolving the dichotomy of island vs.\ loop based devices in the $L\to\infty$ limit. These findings have been successfully employed in the analysis of a recent experiment \cite{vlad}, and will be of future interest in exploring the device's applicability to quantum information processing and observation of Bloch oscillations.

\begin{acknowledgments}
We thank A.\ Kitaev, D.\ Schuster, G.\ Johansson, F.\ Hekking, R.\ Schoelkopf, and S.\ Girvin for valuable discussions.  This research was supported by the NSF under grants DMR-0754613, DMR-032-5580, the NSA through ARO Grant No.\ W911NF-05-01-0365, the Keck foundation, Agence Nationale pour la Recherche under grant ANR07-CEXC-003, and College de France (M.H.D).
\end{acknowledgments}

\end{document}